# MECHANICALLY REINFORCED MgB$_2$ WIRES AND TAPES WITH HIGH TRANSPORT CURRENTS

W. Goldacker, S. I. Schlachter, H. Reiner, S. Zimmer, B. Obst, H. Kiesel, A. Nyilas

*Forschungszentrum Karlsruhe, Institut fuer Technische Physik*
*P.O.Box 3640, D-76021 Karlsruhe, Germany*

## 1 INTRODUCTION

Since the discovery of superconductivity in MgB$_2$ by Akimitsu et al. [1], already all kind of samples, bulk, thin films, wires and tapes were prepared and investigated during a short period [2]. The mechanism of superconductivity is strong electron phonon coupling and can be described by BCS theory. MgB$_2$ therefore belongs to the group of LTC intermetallic systems with an extraordinary high $T_c$ of 39-40 K. Grain boundaries serve as pinning centres and no granularity was observed [3,4]. Several authors reported on powder-in-tube (PIT) wires and tapes with very high transport critical current densities up to about $2 \cdot 10^5$ Acm$^{-2}$ and magnetically measured currents of even $10^6$ Acm$^{-2}$ at 4.2 K and self field [5-8]. A heat treatment of the final conductor with phase decomposition and reformation at temperatures in the range 875-950°C was found to be beneficial to the filament densification and grain connection, the poor sintering of MgB$_2$ and the recovery of strain induced broadened $T_c$ transitions. Treatments at lower temperatures always led to reduced critical currents. Also in cold worked tapes high critical current densities were reported favoured through an efficient material densification from the rolling pressure [9]. Tapes achieved generally significant higher $J_c$ values compared to wires, mostly a consequence of the higher deformation pressure of rollers in comparison to swaging which improves filament densification during rolling. A contribution from deformation texture of the phase was not found, which was checked by X-ray pole figures, and the critical currents which were isotropic for orientation of the tape parallel and perpendicular to the field [10]. For application in devices, especially AC operated devices in energy technique, however, a round wire shape is preferred against a flat tape to achieve low AC losses and to support a layered coil winding. The necessity to improve the transport currents for a competition to NbTi and Nb$_3$Sn needs preparation of *c*-axis textured MgB$_2$ filaments which favours the use of strip, tape or thin film geometries. Investigations on thin films with textured MgB$_2$ [11,12] demonstrated the potential for the achievable critical current densities of $J_c > 10^7$ Acm$^{-2}$ (4.2 K, self field) and had as expected anisotropic superconducting properties with upper critical fields $H_{c2}$ up to 40 Tesla (H || *a, b*). In a quite recently reported thin film sample extraordinary high current densities of $J_c > 10^5$ Acm$^{-2}$ at 4.2 K, 10 T, and an irreversibility field of 21 T (4.2 K) was measured and $H_{c2}$ = 40 T was confirmed [12]. Investigations on single crystals confirmed the anisotropic superconductivity in MgB$_2$ [13].

As sheathed conductors so far predominantly monofilamentary wires and tapes were prepared. Multifilamentary tapes had a significantly reduced current density obviously due to the effect of phase inhomogeneities in the much smaller filament size [14]. Above a certain current density level of about 2·10$^5$ Acm$^{-2}$, which is reached in low fields ($B$ < 4-5 T) or self field at 4.2 K, transport currents are limited due to a thermal instability of the conductors caused by the large filament size, phase impurities, inhomogeneity and especially a non perfect densification of the superconductor. In this range transport currents can become one order of magnitude smaller than calculated from magnetic measurements which give a potential of $J_c = 10^6$ Acm$^{-2}$ (4.2 K, self field) and more. The optimisation and preparation route of the precursor plays a crucial role for reduced secondary phases and improved microstructure and filament density. Since wire and tape preparation involves a heat treatment around the decomposition temperature of MgB$_2$ (875-950°C) the choice of the sheath material in contact with MgB$_2$ is restricted for chemical reasons, due to the reactivity of volatile Mg or the formation of borides with the sheath metals. Ta and Nb tend out to be suitable but form more or less surface reaction layers depending on the precursor route, Fe sheaths react not or only slightly, Ni sheaths very strongly [5-8]. For a mechanical stable wire the thermal expansion of the sheath needs to be larger compared to MgB$_2$ in the temperature interval between 4.2 K and about 1250 K to achieve a stabilising pre-compression of the filament. For compensation of the too small thermal contraction of Nb and Ta, application of a mechanical steel reinforcement is necessary [14,15]. Since Fe sheaths become soft during annealing and have to withstand the Mg pressure, also in this case a Fe/steel sheath improves the wire performance [15]. The disadvantage of such stabilising sheaths are strain effects in the filaments which cause degradations of the superconducting properties quite similar to Nb$_3$Sn wires [15,16]. Studies of $T_c$ under pressure showed the sensitivity of MgB$_2$ on strain with a quite different response on hydrostatic or non-hydrostatic experimental conditions with the occurrence of irreversible $T_c$ degradations in the non hydrostatic case for stresses above 1 GPa as is shown in Fig. 1 (details see ref. [17]) with possible consequences for wires and tapes. Technical wires have, depending on the specific applications, to withstand applied stresses up to 250 MPa and more. The properties of MgB$_2$ in reinforced composites is therefore of most importance for a future technical application and the mayor aspect in this article.

## Crystal structure, thermal expansion and residual strain of MgB$_2$

MgB$_2$ has a simple hexagonal C32 crystal structure (space group P6/mmm), the most common structure among the diborides, which consists of alternating layers of Mg and B.
A strong anisotropy of the thermal expansion and the elastic properties of the *a*- and *c*-axis of the hexagonal crystal lattice (space group P6/mmm) was measured between 4.2 K and RT [18,19]. The thermal expansion in *c*-direction, normal to the sequence of Boron and metal layers is about twice as large as in *a*-direction. This anisotropy is a not unusual property of this class of compounds, but extraordinary strong in $MgB_2$. In dense isotropic material processed at high temperatures, this anisotropic thermal expansion is expected to lead to residual strain load in the grains after cooling. This aspect has not been regarded so far but can be of crucial importance for wire composites, where additional strains are caused through the sheath. An extrapolation of the lattice expansion data of ref. [18] to the phase formation temperature (appr. 900°C) gives a difference of the relative axis elongation between *a* and *c*-axis of the order of 1%. This should lead in fully dense material after cooling to lHe temperature to expected residual lattice strains of the order of 0.5 %.

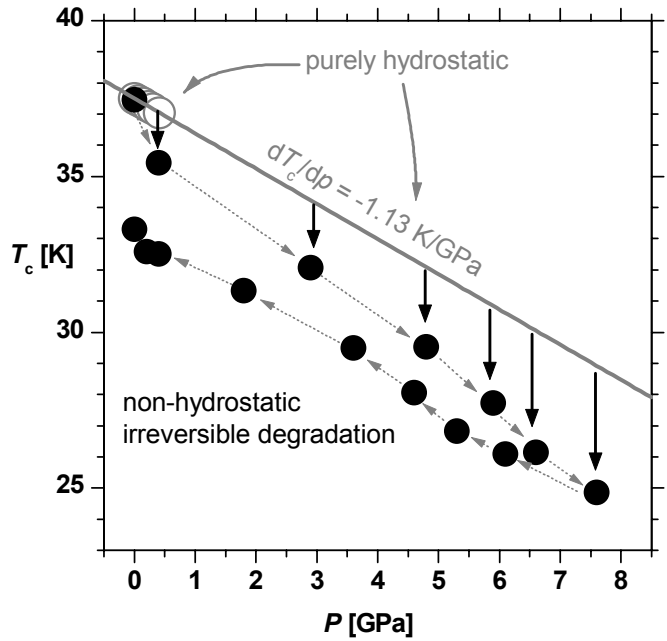

**Fig.1** Critical Temperature $T_c$ of $MgB_2$ with applied pressure $P$. Hydrostatic conditions were performed in a Helium pressure cell, non hydrostatic conditions were given in a diamond anvil cell (see ref. [17]).

Pressure application to $MgB_2$ samples leads to anisotropic strains of the lattice axis due to an anisotropic compressibility [17-19] and degradations of the superconducting properties which was mainly investigated with pressure effects on the critical temperature $T_c(P)$. For $T_c(P)$ a large $T_c$ dependent scattering of the results between 0.2 and 2 K/GPa was reported by different authors [2] obviously correlated to the level of $T_c$, the hydrostatic nature of stress application [17] and the different sample quality. For perfect hydrostatic conditions using He gas and $MgB_2$ powder [18], the largest anisotropy of the pressure induced lattice strains was observed. The highest $T_c$ obviously corresponds to a maximum length of the *c*-axis and changes of the *c*-axis length were analysed to influence strongly in-plane phonons of the B planes ($E_{2g}$ modes) which are strongly coupled to the B sigma bands contributing to the Fermi-level density of states. Studies of substitutional effects in $Mg_{1-x}Al_xB_2$ [20] with the goal to increase $T_c$, showed a sensitive correlation between increasing Al content, decreasing $T_c$ and decreasing length of the *c*-axis. So far all substitutional additives as Mn, Co, Zn, Si, C, Al and others led to a more or less pronounced decrease of $T_c$, indicating that $MgB_2$ is obviously close to a structural instability. No $T_c$ increase was observed so far.

From all results it is well established that structural changes influence sensitively the superconducting properties. These structural changes can result from chemical substitutions and vacancies but also from residual strains and externally applied strains and will have significant consequences for technical conductor composites with mechanically reinforced sheaths.

An obviously first indication of residual strain effects on the superconductivity was observed comparing $MgB_2$ powder and a fully dense HIPped bulk sample [21]. A $T_c$ drop of approximately 0.7 K was observed which could be explained by the expected residual strains in the lattice.

## 2 EXPERIMENTAL

### 2.1 Composition of wires and tapes

For a mechanical stable technical wire a compressive pre-strain in the superconducting core is necessary, which provides a kind of "mechanical reserve" compensating externally applied tensile stresses and strains over a specific range. Therefore not only chemical reasons, the reactivity of the sheath with $MgB_2$, phase purity and microstructure are important features of a conductor design, but also the correct combination of the thermal expansion properties of the sheath components with respect to $MgB_2$ which is responsible for an effective filament pre-compression. Sheath components used in our investigations were Nb, Ta, in combination with Cu and stainless steel (SS) or Fe with and without steel. In table 1 the RT linear thermal expansion coefficients and the specific resistivity are summarised for these materials. For comparison, the data of the Chevrel phase $PbMo_6S_8$ is added, since material combination, conductor concept and the situation for the thermal expansion coefficients was very similar [22,23]. In Fig. 2 the thermal expansion of the composite materials over temperature normalized to $T = 10$ K is shown. For $MgB_2$ the data above room temperature were extrapolated from the crystallographic data below RT (unit cell volume in ref. [18]) using the Einstein equation for a single phonon energy and the cubic root of the unit cell volume data as average value. All materials with a larger thermal expansion than $MgB_2$ can create pre-compression of the filament upon cooling. This is not the case for Nb and Ta, where only a combination with steel can shift the thermal expansion of the composite sheath to higher values than MgB$_2$. The amount of pre-compression of the filament is given through the filling factor, sheath thickness and amount of reinforcement (steel content), the temperature below which all materials become elastic and finally by the elastic properties of the materials, Young's modulus and Poisson ratio. In MgB$_2$ the whole situation becomes more complicated through the expected additional contribution from residual strains caused through the anisotropic thermal contraction of the lattice axis as already mentioned above.

| Material | $MgB_2$ | Ta | Nb | Fe | Ni | Cu | SS | $PbMo_6S_8$ |
|---|---|---|---|---|---|---|---|---|
| $\rho$ [$10^{-8}\Omega m$] at T = 10 K | 0.4 - 16 | 0.1 | 0.1 | 0.024 | 0.006 | 0.002 | - | - |
| $\alpha$ [$10^{-6}$ K$^{-1}$] at T = 298 K | 8.3* | 6.3 | 7.3 | 11.8 | 13.4 | 16.5 | 18 | 9.4 |

**Tab.1** Specific resistivities $\rho$ and linear thermal expansion coefficients $\alpha$ for $MgB_2$ and sheath materials. $\alpha$ from $MgB_2$ was calculated from the expansion of the unit cell volume in ref. [18], the values for the Chevrel phases are added for comparison. SS = stainless steel.

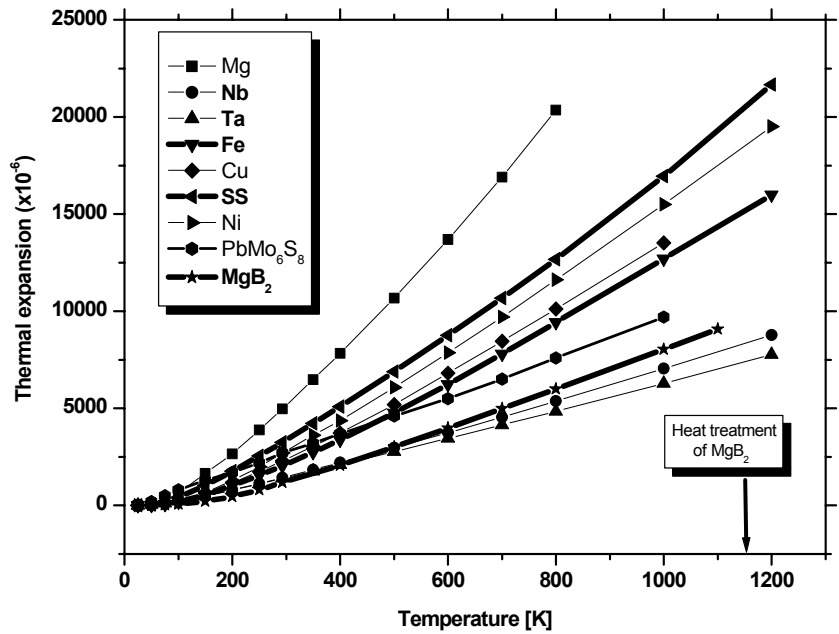

**Fig.2** Thermal expansion of MgB$_2$ and different sheath materials normalised to $T$ = 10 K. For comparison the values for Chevrel phase PbMo$_6$S$_8$ are added. SS = stainless steel

| Sample | Filament | Fe or Nb | Cu | Steel |
|---|---|---|---|---|
| MgB$_2$/Fe | 49 | 51 | - | - |
| MgB$_2$/Fe/SS25 | 30 | 33 | - | 37 |
| MgB$_2$/Fe/SS35 | 22 | 24 | - | 54 |
| MgB$_2$/Fe/SS45 | 16 | 18 | - | 66 |
| MgB$_2$/Nb/Cu/SS25 | 28 | 18 | 21 | 33 |
| MgB$_2$/Nb/Cu/SS35 | 20 | 13.5 | 15.5 | 51 |
| MgB$_2$/Nb/Cu/SS45 | 14 | 10 | 12 | 64 |

**Tab.2** Composition of ex-situ MgB$_2$-wires with Fe and stainless steel (SS) reinforced sheath. Material contents were estimated from micrographs of the wire cross sections.

## 2.2 Wire, tape preparation

MgB$_2$ has poor sinter properties due to volatile Mg and has as main secondary phase MgO. Therefore, two different precursors were applied in our wires and tapes. In the first concept, commercial MgB$_2$ powders (Alfa Aesar) with about 2% MgO secondary phase (estimated from X-ray spectrum) - obviously varying with production batch - were used (ex-situ-samples). In the second route, powder mixtures of Magnesium (< 300 mesh, 99 % purity) and Boron (< 200 mesh, 99 % purity) were used (in-situ-samples) as a promising way reducing the oxygen ingot, which forms MgO, and to improve the filament densification due to the melting of Mg above 650°C. The precursor powders were mixed for several hours and were not especially ground before filling the sheath tubes. The filling density was in the range 50 to 70% depending on the precursor.

Ta and Nb tubes were applied as sheath (10mm diameter, 1mm wall thickness) in combination with a Cu tube of 12 mm outer diameter (to allow wire drawing) or alternatively Fe tubes (99.8% Fe). Deformation was made by swaging and drawing to 1.5-1.8 mm wire diameter. At this stage, the wire was introduced in a stainless steel (SS) tube of 2.5 mm outer diameter, 0.25, 0.35 or 0.45 mm wall thickness, and deformed to the final wire diameters between 0.7 and 1.58 mm diameter (see table.2). Steel reinforced wires were produced with Nb/Cu/SS and Fe/SS sheaths (Fig. 3,4). The necessary amount of stainless steel was estimated roughly from the experience with Chevrel phase wires in refs. [22,23] and varied between 33 and 66 %. Wire lengths of 5 - 18 m are routinely produced. All wires had quite regular cross sections. The final heat treatment was performed at 875 - 950°C, which is above the start of phase decomposition, followed by a cooling ramp with 10 – 20 °C/hr to 800 °C for phase recovery with final furnace cooling to RT. The annealing atmosphere was Ar/H$_2$(5%) to avoid oxidation of the steel sheath. For the first sample batch identical heat treatments were applied

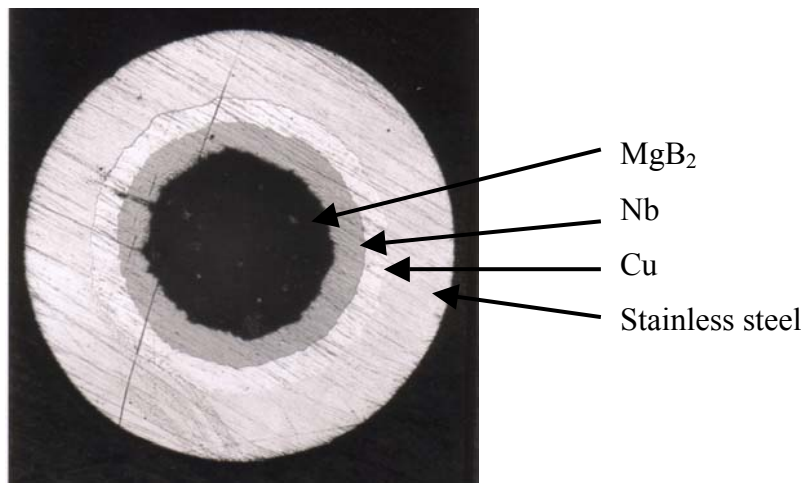

**Fig.3** Cross section of a reinforced MgB$_2$/Nb/Cu/SS- wire with 1.18 mm diameter after the annealing process.

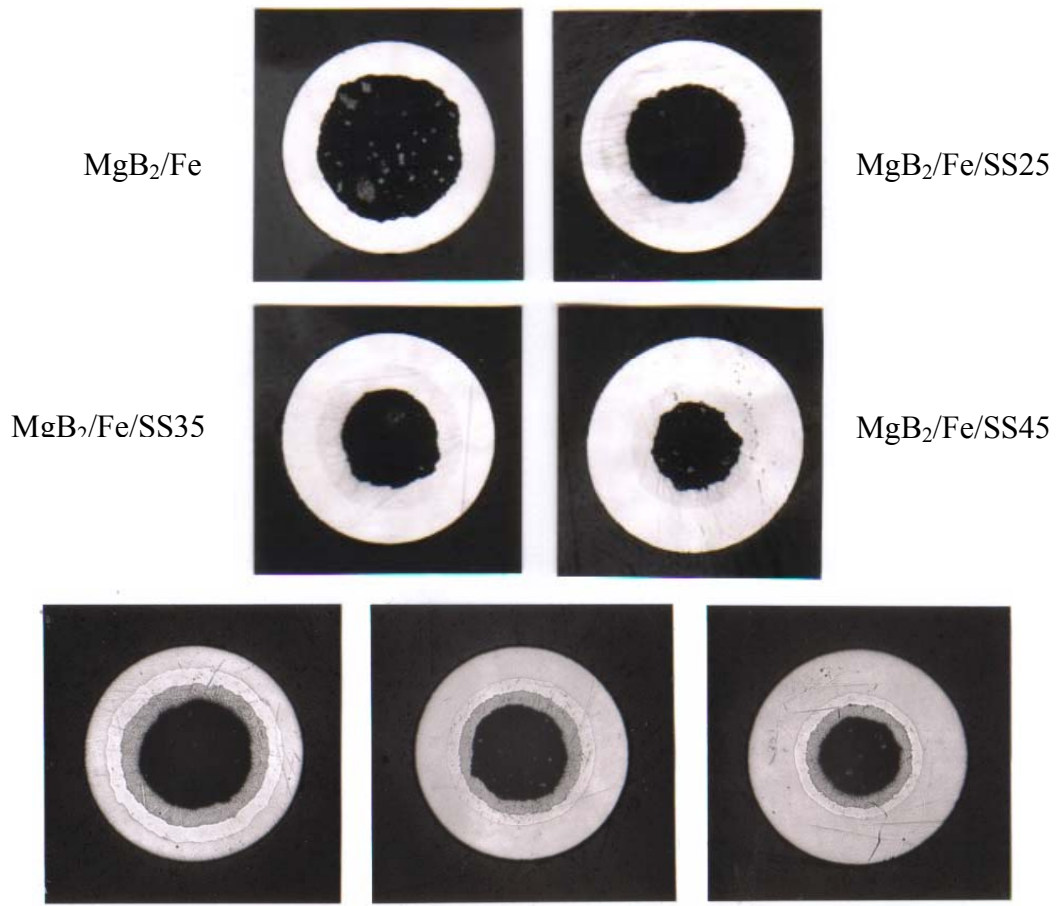

**Fig.4** Cross sections of reinforced ex-situ MgB$_2$-wires (see Tab.2). In the lower part from left to right MgB$_2$/Nb/Cu/SS25, MgB$_2$/Nb/Cu/SS35 and MgB$_2$/Nb/Cu/SS45 -wire

to ex-situ and in-situ wires. This preparation scheme was not the optimum for the specific in-situ Mg/B precursor route. A systematic optimisation of the annealing treatment is not finished so far, but meanwhile much progress was achieved applying an additional heat treatment during deformation. This pre-reaction of MgB$_2$ avoids deformation problems of Mg and B powders at small wire diameters and leads to an improved material mixture of reacted and non reacted material during the final deformation treatment.

MgB$_2$ tapes were produced from ex-situ Fe-sheathed wires at diameters of 1.5 – 2 mm, using parallel hard metal rollers of 95 mm diameter and applying 10% reductions steps. Tapes were only investigated for Fe sheaths so far to establish the systematic difference to the wire concept. A row of tapes with different aspect ratios of the cross sections were prepared from one single round wire sample as shown in Fig. 5. The round wire and a turk head rolled square wire serve as reference, all samples were annealed together.

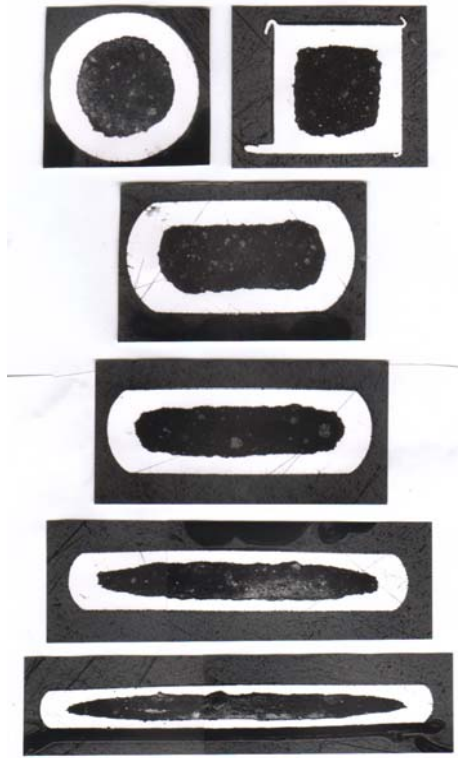

**Fig.5** Cross sections of ex-situ MgB2/Fe wires and tapes given in same magnification. The square turk head rolled wire (top right) has a dimension of 1.1 x 1.1 mm.

## 2.3 Characterisation methods

The precursor powders were analysed by X ray diffraction, the particle size distribution by means of a Beckmann-Coulter LS230 particle size analyser. Microstructures in the filament cross sections were investigated with SEM and EDX. Phase formation was studied by means of a differential scanning calorimeter (Netzsch Jupiter 449 DTA/TG/DSC).

$T_c$ values were measured inductively, the methods for applying pressure are given in ref. [17] in detail. Transport critical current measurements were determined by means of conventional 4-probe method in perpendicular background fields up to 12 T in lHe or in a flow cryostate at temperatures between lHe and $T_c$. In a special miniature strain rig for high fields [24], axial stress up to about 700 MPa could be applied to a short wire sample of 40 mm length by means of a stepper motor. Applied stresses were measured with a piezo load cell close to the sample and strains were measured directly over a 14 mm section of the wire via an attached DMS precision strain gage clip. Stress-strain experiments were confirmed with measurements on a high precision strain rig on 180 mm long samples in the background field of a 13 T split coil magnet. Transport currents were measured for each strain value loading the sample with a fast current ramp of typically 0.2 sec scan time. Standard criterion for the determination of $I_c$ was 1 µV/cm.

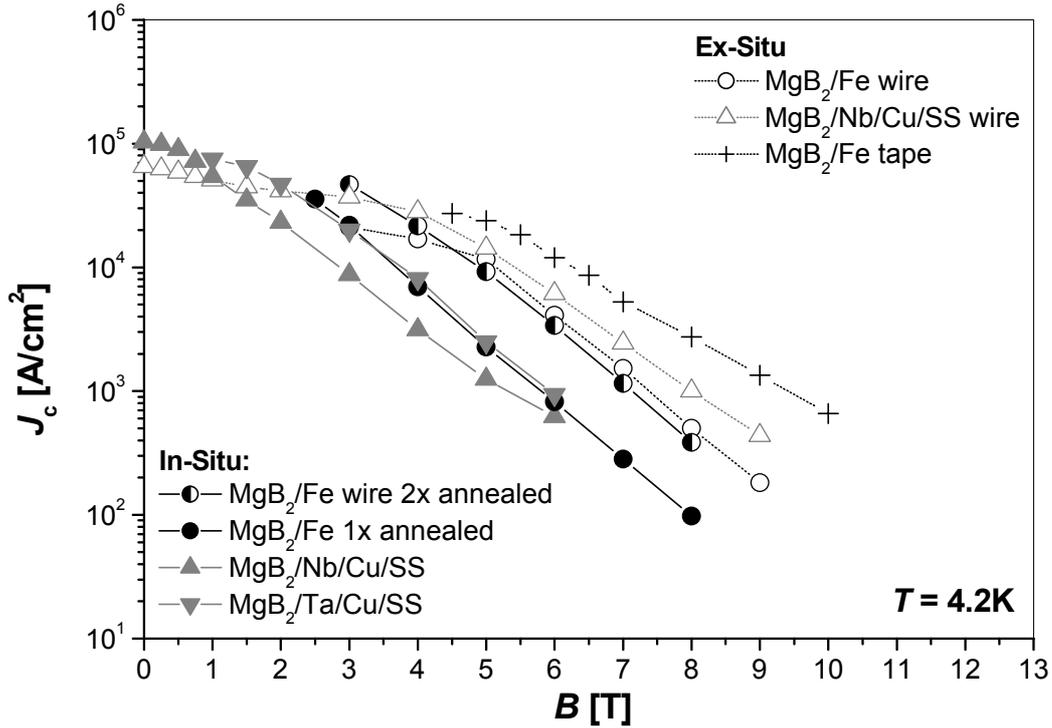

**Fig.6** Critical current density of different $MgB_2$ wires and a tape with different sheath combinations and in-situ and ex-situ precursor route. Explanations see text

## 3 RESULTS

### 3.1 Transport critical currents of in-situ and ex-situ wires

For in-situ and ex-situ $MgB_2$-wires, prepared with identical deformation and heat treatment, a quite different behaviour of the transport currents with field was observed as is given in Fig. 6. Ex-situ wires from pre-reacted powders reached 65-100 kAcm$^{-2}$ at 4.2 K, 0 T and possess still relative high transport currents at high fields (see table 2), which corresponds to an irreversibility field of 10-11 T, estimated from Kramer plots. SEM investigations of the microstructure in the filaments indicate a quite homogeneous phase distribution but a quite broad distribution of grain sizes in the major part of the cross section. In-situ wires received higher current densities of $10^5$ Acm$^{-2}$ (4.2 K) at small fields or at self field, but with a much larger degradation of $J_c$ in increasing fields, and $H^*$ extrapolates to only 7-9 T. The material of the first wall, Ta, Nb and Fe, plays an important role for the selection of the precursor route due to the different influence of chemical reactions between filament and sheath. This different behaviour influences also the microstructure as a consequence of the specific phase formation mechanism. The in-situ reaction of the phase begins with melting of Mg above 650°C. At temperatures above 850°C Boron grains react from the surface to the inner part forming $MgB_2$. Therefore, the grain size of the Boron particles determines significantly the microstructure. This leads to much larger $MgB_2$ grains and a more inhomogeneous microstructure compared to ex-situ wires [25]. Some additional formation of borides with the sheath materials, as $NbB_2$, can disturb the filament stoichiometry. Since flux pinning obviously occurs at grain boundaries as pinning centres, the microstructure has a crucial influence on the final superconducting properties and is one reason for the observed variation of either, the superconducting currents at higher fields and the irreversibility fields. For a comparison with the wires, $J_c(B)$ of the best $MgB_2$/Fe tape is also given in Fig. 6. An explanation for the significantly higher current density level is given in the next section.

Generally at fields below 3-4 T the slope of the $J_c(B)$ curves narrows in contrast to magnetic measurements which show a mostly linear characteristics of $\ln J_c(B)$ [25]. The reason is a poor thermal stabilisation of the wires if locally $I_c$ is reached. A non sufficient densification of the filament material and the presence of inclusions of secondary phases as MgO are the most probable reasons. In this field regime many samples heat up very fast to the normal conducting state above $I_c$ and even burn through. The $U(I)$ characteristics of the $I_c$ transitions show the sudden quench of superconductivity. The Cu stabilisation in the Nb/Cu/SS sheaths didn't improve the low field critical current transition but avoided widely the burn trough effects. Therefore, an improved densification, a better phase homogeneity and microstructure and especially a multifilamentary arrangement of the core is required to realize thermally stable wires with carry the full gain of transport currents.

For in-situ wires a significant improvement was achieved, varying the stoichiometry to $Mg_{0.9}$ and applying two reaction heat treatments (RHT), one during deformation, the second in the final wire. The effect of the intermediate heat treatment is the formation of $MgB_2$ which is ground during the further deformation due to its brittleness. This improves the filament homogenisation as a favourable condition for the final reaction treatment. Also an improved filament densification was observed in SEM pictures. With this modified preparation method, in-situ wires reached already the current carrying capability of the ex-situ wires (see Fig. 6).

## 3.2 Ex-situ $MgB_2$-tapes

Tape conductors are less favourable for low AC losses due to geometrical reasons and are also less suited to perform layered coils, but can obviously carry significantly higher transport currents. A series of tapes was prepared and investigated to look systematically for the reasons. In Fig. 4 the cross section of a round wire, a rolled square wire and tapes with different aspect ratios are shown. All conductors are made from one piece and thermally treated together. In Fig. 7 the critical transport current densities with field show a systematic increase with tape character of the conductor. Also the slope of the $J_c(B)$ dependence changes indicating an increase of the irreversibility field from about 10-11 T in the round wire to about 14-15 T in the tape. A comparison of the current densities at 6 T plotted against the aspect ratio of the conductor cross section, shown in Fig. 8, gives an increase by a factor of 3 with a tendency to even higher values but also towards some limit. A speculating polynomial extrapolation to much higher aspect ratios, leads to a limit of about 40-50 kAcm$^{-2}$ for presumably fully dense samples, which is approximately the current level found by magnetic measurements in the best tapes, representing the intra-grain current level. Therefore, the $J_c$ increase can be explained and understood as the effect from improved filament densification caused by the rolling pressure. For excluding an influence of texture effects, X-ray pole figure patterns on filament material, separated from the tapes, were performed and a possible anisotropy of the critical transport currents was checked measuring $I_c$ for $B$ parallel and perpendicular to the tape surface. Neither a *c*-axis texture nor different critical current values were found, supporting our interpretation.

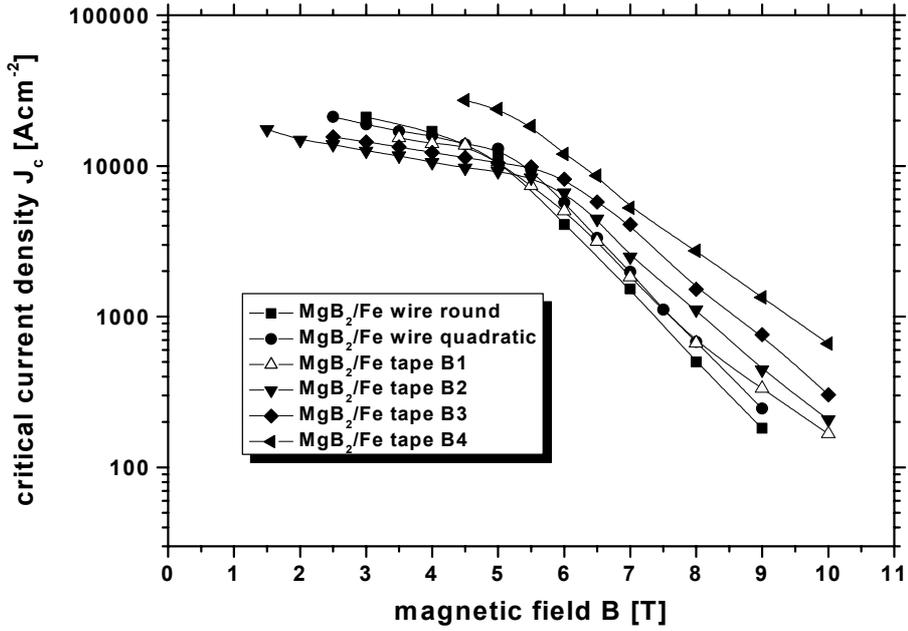

**Fig.7** Critical current densities with magnetic field for ex-situ $MgB_2$/Fe wires and tapes with different aspect ratios as given in Fig. 5

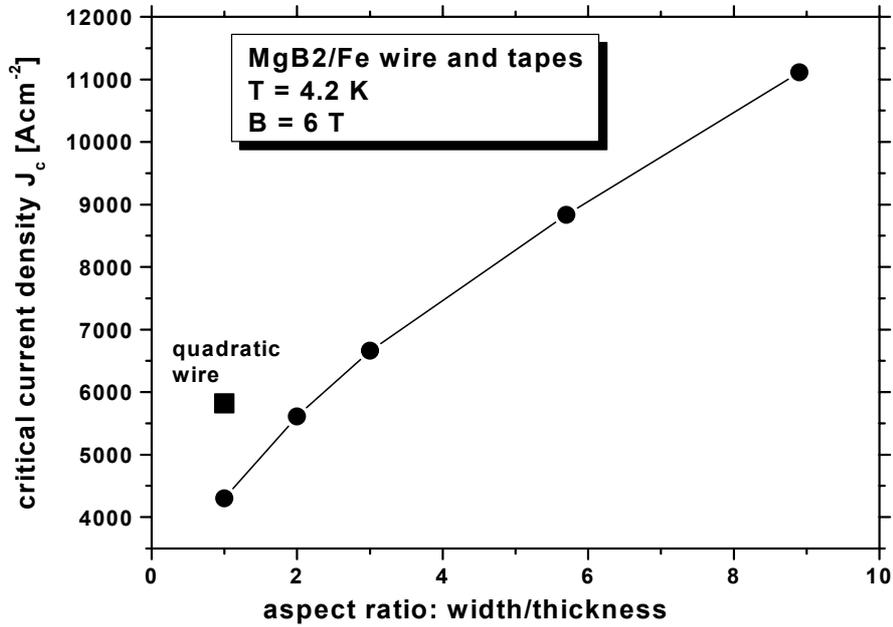

**Fig. 8** $J_c$ vs. aspect ratio at $B = 6$ T for $MgB_2$/wires and tapes from Fig. 7 (see also Fig. 5)

The observed increase of the critical fields in the tapes can also be understood. The $MgB_2$ precursor particle sizes distribute from sub-micron size to >100 microns (agglomerates). With improved densification of the filaments upon rolling the sample volume contributing to the current percolation path obviously increases. Smaller particles in the filament need higher pressure for densification and to make an effective intergrain contact. The increase of rolling pressure with smaller tape diameter therefore favours the densification of small grained sections and improves the transport current carrying sample volume. Since flux pinning is grain size related, the contribution from small grains leads to improved high field currents and higher irreversibility fields.

### 3.3 Effect of mechanical reinforcement

The addition of different steel amounts as mechanical reinforcements in the sheath (see tab. 2 and chapter 2.2) was performed to study the effects of pre-stress in the filaments and the consequences for the transport currents. Systematic investigations were performed on ex-situ wires. In the $J_c(B)$ graphs for both sheath concepts, Nb/Cu/SS and Fe/SS, raised steel contents led to field dependant reduced currents, especially at high fields as is shown in Fig. 9 and Fig. 10. This behaviour is well known from $Nb_3Sn$ and Chevrel wires and can be attributed to pre-stress induced degradations of the superconducting properties, the critical current and the irreversibility field. For Nb/Cu/SS sheaths, the averaged thermal expansion coefficient is smaller in comparison to Fe/SS sheaths, therefore also the degradation of the transport currents for comparable steel content occurs to be smaller.

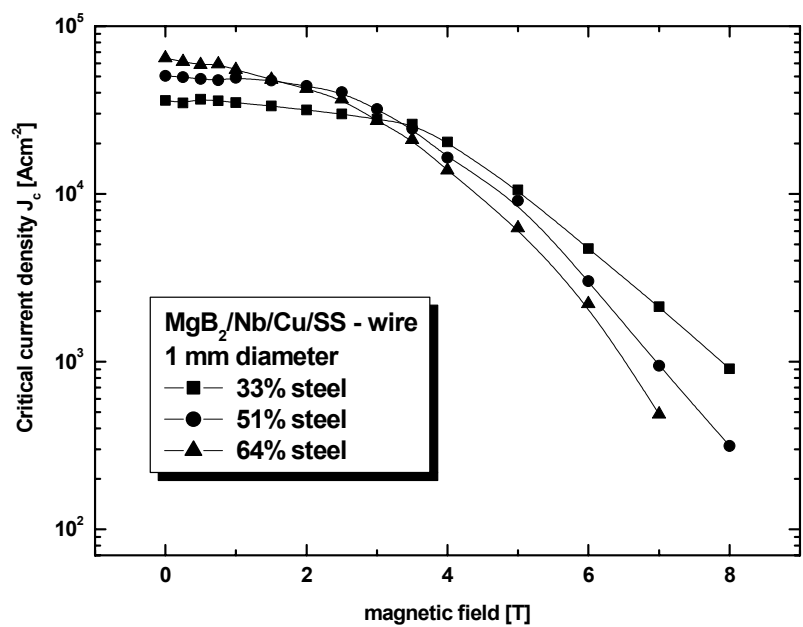

**Fig.9** Critical current densities with field for ex-situ $MgB_2$/Nb/Cu/SS wires with different content of steel reinforcement (see Fig. 4)

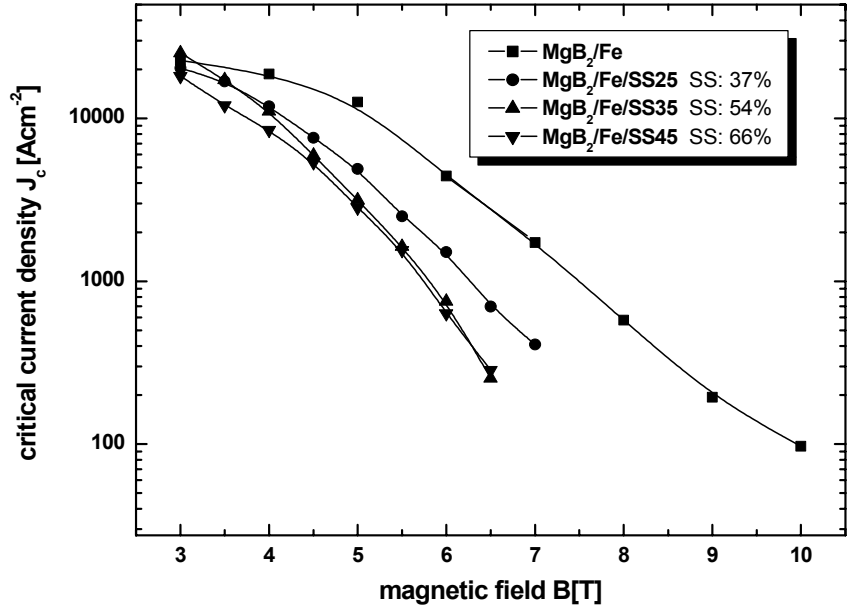

**Fig.10** Critical current densities versus field of ex-situ $MgB_2$/Fe and $MgB_2$/Fe/SS wires with different steel contents (see Fig. 4)

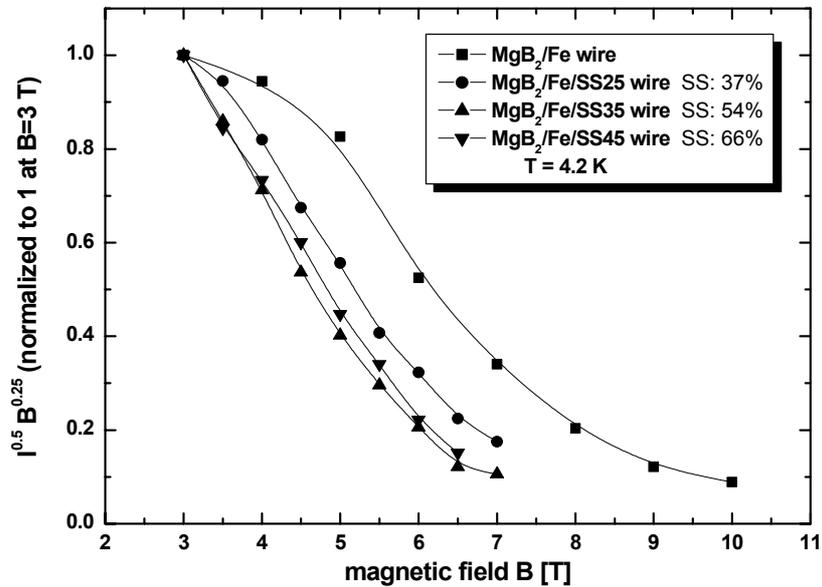

**Fig.11** Kramer plot for the $J_c(B)$ graphs of Fig. 10

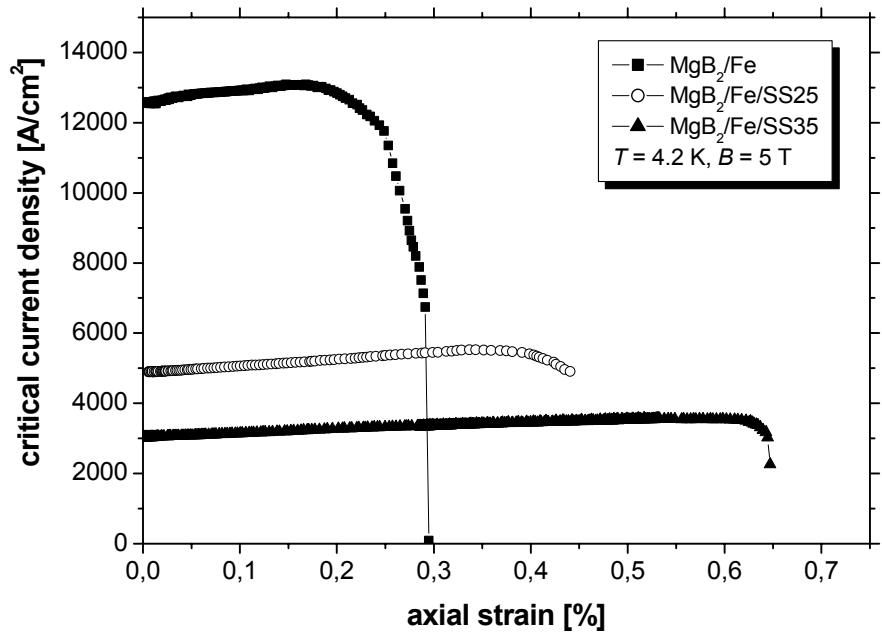

**Fig.12** Critical current density versus applied axial tensile strain for $MgB_2/Fe$ and $MgB_2/Fe/SS$ wires in a background field of 5 T. The wire diameter was 1 mm.

In $MgB_2/Nb/Cu/SS$ wires (Fig. 8) above 3 Tesla the currents reduce with steel content, below 3 Tesla they increase significantly. The improved currents at low fields are obviously from a better filament densification with the radial compressive stresses from the sheath. The pre-strain effect on the irreversibility fields can better be seen in the Kramer plot representation (Fig. 11), the extrapolated values decrease from 11 to about 8 Tesla.

In $MgB_2/Fe/SS$ wires (Fig. 10) the addition of steel causes much stronger current degradations than in $MgB_2/Nb/Cu/SS$ wires since already a Fe sheath makes some pre-compression of the filament. For the highest steel contents the effect saturates. Consequently also the change of $H^*$ is larger from about 13 Tesla down to about 7-8 Tesla.

### 3.4 Critical currents with applied axial tensile strain

Stress-strain experiments with simultaneously measured change of the critical currents were performed for the $MgB_2/Fe/SS$ wires in a background field of 5 T. In these experiments the compressive axial pre-strain of the filaments is compensated applying axial tensile strains. A recovery of the strain induced $I_c$ degradation is expected. The amount of the strain induced degradation of the critical current density with applied steel reinforcement can be obtained from the $J_c(B)$ graphs in Fig. 10. For 33% steel addition to the sheath $J_c$ reduces to 40% with respect to the value of the non reinforced Fe sheathed wire, for 51% steel content to 25% of $J_c$.

For all three investigated wires an increase of the critical currents with applied strain was measured as given in Fig. 12. In the regime of increasing critical currents all current changes were completely reversible. $I_c$ degradations observed above some sample specific strain limits, 0.2% for the Fe sheath, 0.4 % for the Fe/SS25 sheath and 0.6% for the Fe/SS35 sheath, however were irreversible and have to be attributed to a damage of the superconductor caused by formation of cracks in the filament.

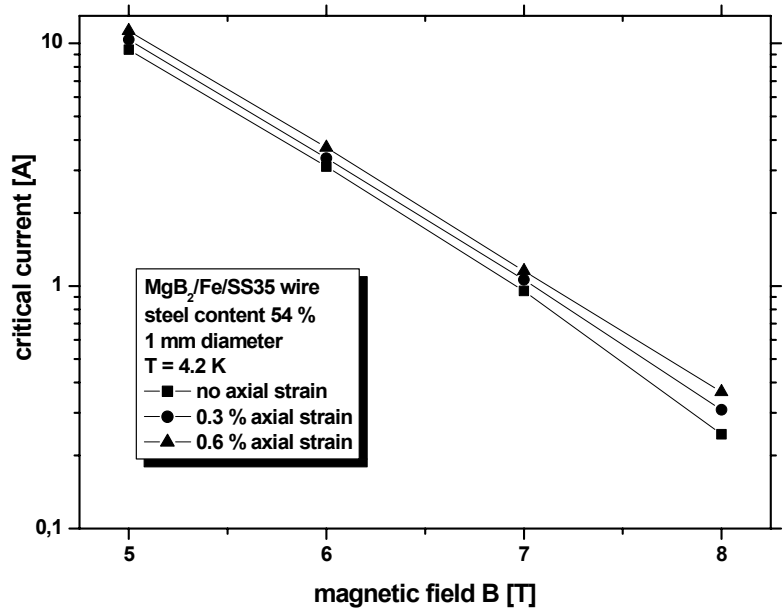

**Fig.13** Field dependence of the critical currents of an ex-situ $MgB_2$/Fe/SS wire with 51% steel content for axial strains of 0%, 0.3% and 0.6%

The observed relative increase of the transport currents was slightly larger for the steel reinforced wires (see Fig. 13), but only a small fraction of the $J_c$ degradation with respect to the Fe-sheathed wire was recovered. Quite different effects obviously play a role. First, a still high hydrostatic stress component in the strained samples exists, which scales with conductor reinforcement. 3-dimensional calculations of the stress and strain state during an $I_c$ vs. applied strain experiment for the quite similar example of Chevrel phase wires showed [26], that with a compensation of the axial compressive pre-strain component even an increase of the radial compressive strain components correlates. This explains the non complete recovery of the current degradation for the reinforced wires compared to the non reinforced, which is a quite common observation for the axial applied strain experiment. But this cannot explain the whole effect observed in the reinforced $MgB_2$ wires, since the fraction of the recovered current is quite small. This lack of current recovery must be attributed to a damage of the superconductor as observed in the $T_c(P)$ data (see Fig.1). The shear components of the pre-strains obviously lead to irreversible degradations of the superconducting properties, in particular a reduced irreversibility field and irreversibly reduced current densities at higher fields. If this effect is an intrinsic property of $MgB_2$ cannot be decided so far due to the present imperfect sample quality. The microstructure may play an important role since crack formation occurs preferred at grain boundaries and a preferred disconnection of small grains may be an explanation for the observed effects.

For the strain experiment of wire $MgB_2$/Fe/SS35, $J_c(B)$ dependences were measured at applied strains of 0%, 0.3% and 0.6% to check the influence of applied strains on the irreversibility fields. The results are given in Fig. 13. Only a very small shift to higher current values is observed, corresponding to only a slight increase of the irreversibility field with released pre-strain. This results confirm too, that the strain induced degradation of $H^*$ is due to irreversible damage of the sample.

## 4  DISCUSSION AND CONCLUSIONS

In this contribution, different concepts for mechanically reinforced $MgB_2$ wires with high transport currents were presented and characterised. The achieved critical transport currents depend strongly on the conductor concept, the choice of the precursor route and the applied treatment method. The future increase of the current carrying capability of $MgB_2$ wires and tapes and therefore their chance for a technical application depends strongly on an improvement of the material quality, homogeneity and the wire composite design. An optimised small grained microstructure in the filaments requires new or improved precursor preparation techniques. The anisotropic superconducting properties of $MgB_2$ ask for a textured superconducting core in future wires or tapes. $MgB_2$ wires show, depending on the sheath composition, strong strain induced effects on the superconducting properties. The observation that steel reinforcements of wires lead to a drastic pre-strain induced current degradation and that applied strain compensating the pre-strain recovers $J_c$ only by a small amount, is an indication of a more complex situation for $MgB_2$ in a wire composite. Stresses in wire cores are known to be non hydrostatic which favours irreversible effects and defects in $MgB_2$. Expected residual strains due to anisotropic mechanical properties of the crystallites in dense samples complicate the boundary conditions in addition. A filament pre-compression however is absolutely necessary for technical wires to compensate externally applied strains occurring during coil winding or at operation in coils from Lorentz forces. This pre-stress is also very favourable balancing the pressure of volatile Mg during the annealing process of the wire and supporting the filament densification during phase reformation. For optimised conductor designs the strain sensitivity of $MgB_2$, however, requires a very well balanced amount of reinforcement to achieve an optimised compromise between degradation of the superconducting properties and the advantages from the reinforcement. However, all results indicate a wide gain of further possible improvements of all conductor properties, especially the current carrying capability, and raise the chance for a future successful competition with commercial LTC superconducting wires for selected applications.